# Spectral and spatial shaping of Smith Purcell Radiation

Roei Remez[1]*, Niv Shapira[1], Charles Roques-Carmes[2], Romain Tirole[2,3], Yi Yang[2], Yossi Lereah[1], Marin Soljačić[2], Ido Kaminer[2,4] and Ady Arie[1]
[1]School of Electrical Engineering, Fleischman Faculty of Engineering, Tel Aviv University, Tel Aviv, Israel.
[2]Research Lab of Electronics, MIT, Cambridge, MA 02139, USA
[3]Department of Physics, Faculty of Natural Sciences, Imperial College London, London SW7 2AZ, UK
[4]Department of Electrical Engineering, Technion – Israel Institute of Technology, Haifa 32000, Israel
*Correspondence to: roei.remez@gmail.com.

## Abstract
The Smith Purcell effect, observed when an electron beam passes in the vicinity of a periodic structure, is a promising platform for the generation of electromagnetic radiation in previously-unreachable spectral ranges. However, most of the studies of this radiation were performed on simple periodic gratings, whose radiation spectrum exhibits a single peak and its higher harmonics predicted by a well-established dispersion relation. Here, we propose a method to shape the spatial and spectral far-field distribution of the radiation using complex periodic and aperiodic gratings. We show, theoretically and experimentally, that engineering multiple peak spectra with controlled widths located at desired wavelengths is achievable using Smith-Purcell radiation. Our method opens the way to free-electron driven sources with tailored angular and spectral response, and gives rise to focusing functionality for spectral ranges where lenses are unavailable or inefficient.

## A. Introduction

Ever since its first observation in 1953 [1], Smith Purcell Radiation (SPR) has been studied extensively, mainly due to its potential to generate light at hard-to-reach electromagnetic frequencies, such as the far-infrared, terahertz, and x-ray ranges [2–5]. This radiation, which originates from the collective oscillations of charges induced by free charged particles above a periodically patterned surface, is characterized by a specific dispersion relation and emission at every angle in the plane normal to the surface.
The community has explored various theories to account for the physics of this phenomenon in metallic gratings [6–9] and to increase the efficiency of the radiation [2,3,10,11]. Even more advanced theories revealing quantum features of this effect [12,13] assumed simple periodic structures, giving a momentum $n\frac{2\pi}{\Lambda}\vec{x}$ to the incoming electron (where $n$ and $\Lambda$ are the diffraction order and grating period, respectively). Recently, the study of SPR was extended beyond the simple periodic structure, into aperiodic arrays [14], disordered plasmonic arrays [15], and Babinet metasurfaces for manipulating the SPR polarization [16]. However, to the best of our knowledge, no attempt has been made to shape the angular spectrum and frequency spectrum of SPR.
In this Letter, we develop a systematic method for shaping the SPR's angular and frequency spectrum using complex periodic and aperiodic gratings. First, we show theoretically and experimentally that the single peak of the conventional SPR far-field distribution can be split into a desired set of peaks by properly designing a complex periodic grating, and demonstrate twin-peak and triple-peak shaping. We then discuss the advantages and limitations of the method and study the effect of the interaction length on the radiation, showing that it can be used to smoothen the spectral or angular response for a desired outcome. We conclude with a theoretical suggestion corroborated by a set of simulations for a Smith-Purcell cylindrical lens.

### B. A theoretical analysis of Smith-Purcell radiation in complex periodic gratings

We start with writing the expression of the field emitted from an infinite metallic periodic grating, positioned at $(x, y)$ plane with the rulings parallel to the $y$ direction, and where the electron is passing close to the grating with velocity $v_0\hat{x}$, as proposed by Van Den Berg [6]:

$$E_x(x,y,z,t) = \frac{1}{2\pi^2} Re\left[\int_0^{\infty} d\omega \int_{-\infty}^{\infty} d\beta \sum_{n=-\infty}^{\infty} \epsilon_{xn}^{r}(\beta,\omega) e^{i\alpha_n x + i\beta y + i\gamma_n z - i\omega t}\right] \quad (1)$$

where $\alpha_n = \frac{\omega}{v_0} + \frac{2\pi n}{\Lambda} = k_0 \sin\phi \sin\theta_n$ , $\beta = k_0 \cos\phi$ , $\gamma_n = \sqrt{k_0^2 - \beta^2 - \alpha_n^2}, k_0 = \frac{\omega}{c} = \frac{2\pi}{\lambda_0}$, $\lambda_0$ and $\omega$ are the vacuum wavelength of light and its angular frequency, $c$ is the velocity of light, $\phi$ is the angle of the wave vector with y axis and $\theta_n$ is the angle of the projected $k$ vector on the $(x, z)$ plane with $z$ axis. The discrete summation over $n$ originates from the different diffraction orders of the grating, with corresponding spatial frequency components $\alpha_n$ and $\gamma_n$. Although the summation is on all values of $n$, we note that for high diffraction orders $\gamma_n$ becomes imaginary and the wave is therefore evanescent. From the equation for $\alpha_n$ we can formulate the dispersion equation:

$$\lambda_0 = \frac{\Lambda}{n}\left(\sin\phi \sin\theta_n - \frac{c}{v_0}\right) \quad (2)$$

Which is equivalent to the dispersion relation from the first paper by Smith and Purcell [1]. Eq. (1) represents decomposition into plane waves in direction $(\theta_n, \phi)$.

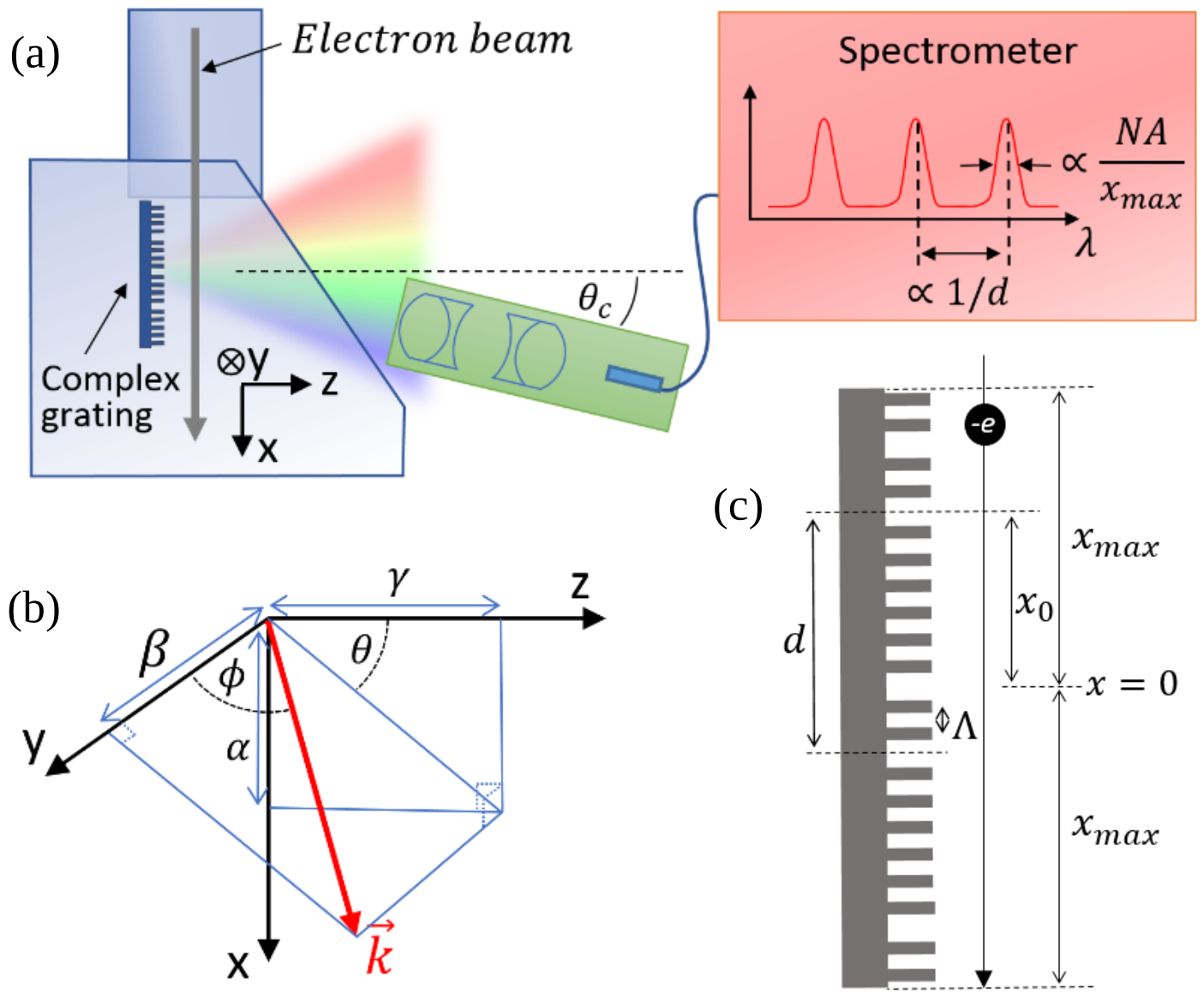

**Figure 1 – Experimental setup.** (a) The special grating was positioned inside the viewing chamber of the transmission electron microscope, while the output radiation is collected using lenses and a spectrometer located outside the microscope. Inset to (a) – the spectral response of the first diffraction order of the radiation exhibit multiple peaks with spacing and width inversely proportional to the dislocation period and the interaction length, respectively. (b) – Definition of angles and momentum components of the emitted light. $\phi$ is the angle between the wave-vector of the emitted light and the y axis, and $\theta$ is the angle between the projection of the $\vec{k}$ vector on the xz plane and the z axis. (c) Schematic drawing of a complex periodic grating, showing the supercell of period d created by half-period spacings on an otherwise constant period Λ.

To exemplify our method, we limit our derivation for the case of constant duty cycle (ridge to period ratio) of 50%, which results with phase-only changes for the electric field. We can define a function $s(x)$ that expresses the structuring of the grating (the results below can be directly generalized to any duty cycle and even arbitrary grating shapes). A shift of $s(x)\Lambda$ (where $0 \leq s(x) \leq 1$) will result in phase shift $\Delta\psi$ for the electric field at *z=0*, given by:

$$\Delta\psi = 2\pi s(x) n$$

where $n$ is the SPR diffraction order (See [17] for Huygens construction of this expression). Here, the function $s(x)$ is assumed periodic of period $d$, however as we discuss in section D this does not have to be the case, and some interesting results can be achieved with non-periodic variations of the grating. The angular far-field distribution can be derived taking into account the structuring of the grating:

$$\eta_x^r(\alpha,\beta,\omega) = 2x_{\max} \sum_n \epsilon_{xn}^r(\beta,\omega) \sum_m a_{mn} e^{-i\left(\alpha - \frac{2\pi m}{d}\right)x_0} \operatorname{sinc}\left(\left(\alpha - \frac{2\pi m}{d} - \alpha_n\right) x_{\max}\right) \quad (3)$$

where $2x_{\max}$ is the interaction length of the electron with the grating, $x_0$ is the position of the interaction area with respect to an arbitrary point in the grating, and $a_{mn} = \frac{1}{d}\int_d e^{i\Delta\psi(x)} e^{i2\pi m x/d} dx$ are the phase Fourier coefficients [17]. If the optical collecting system has a finite aperture $\xi$, centered around angular coordinates $(\theta,\phi) = (\theta_c, \pi/2)$, the measured power for each angular frequency $\omega$ will be proportional to the integration of the intensity of that frequency over the collecting aperture:

$$P(\omega) \propto \iint_\xi \left|\eta_y^r(\alpha,\beta,\omega)\right|^2 d\alpha\, d\beta \qquad (4)$$

For small collecting aperture with angular diameter $A < \frac{2\pi}{dk_0}$ and interaction length larger than the period of the variation, $2x_{\max} > d$, Eq. (3) and (4) give a spectrum with a series of separated peaks. The peaks are centered at wavelengths $\lambda_{mn}$ yielding (for $\phi = \pi/2$):

$$\lambda_{mn} = \frac{sin\theta_c - \frac{c}{v_0}}{\frac{n}{\Lambda} + \frac{m}{d}} \qquad (5)$$

The expression is identical to the ordinary SPR equation (Eq. (2)) for the zero-th order *m=0*, but for $m \neq 0$, additional peaks appear in different locations which are determined by the variation periodicity *d*. Interestingly, the zero-th order amplitude $a_{0n}$ can be canceled, eliminating the original SPR peak, by adequately designing the structuring function $s(x)$ so the phase $e^{i\Delta\psi}$ averages to zero (as in Fig. 2). The width of the peaks is set by the interaction length $2x_{\max}$ and by the aperture of the collection system. Increasing the aperture A, or decreasing the interaction length $2x_{\max}$ (as in Fig. 3) widens the spectral peaks while their central wavelength remains constant, until the peaks become inseparable. Since it is assumed that each electron has a different value of $x_0$, the radiation from different electrons is summed incoherently with equal distribution of $x_0$ between 0 and d.

### C. Spectral shaping of SPR

Fig. 2 demonstrates the modification of the first diffraction order (*n*=-1) of SPR using structures with a half-period shift every M/2 periods of the grating, therefore $d = (M+1)\Lambda$. For example, see Fig. 2(a) for s function when M=16 ($\Lambda = 400nm$ for all gratings). In this case, the two dominant Fourier coefficients in Eq. 3 are $a_1$ and $a_{-1}$, as shown in Fig. 2(b), where we have omitted the n subscript $a_m \equiv a_{m,n=-1}$. Importantly, this nulls the zero-th order Fourier coefficient $a_0$, which defines the amplitude of the original SPR diffraction expression (Eq. 2). See [17] for the technical details of the experiment.

The experimental and theoretical spectral responses for different M values are given in Fig. 2(c-e). The twin-peak spectral shape is clearly visible for M = 16 (Fig. 2c). When M increases, the two peaks get closer to one another, until for M = 64 (Fig. 2e) they are no longer separated, because of the limited interaction distance that depicts the number of unit cells (~30) that the electron is interacting with. By optimizing the first of the measured width to simulation, we estimate the value for the interaction length to be 12 microns. Using this value alone, our theory shows a very good fit to all spectra (red curves in Figs. 2(c-e)). Figs. 2(c-e) where taken for the same value of $\theta_c$ which we measure to be $-7.7 \pm 1.5\ deg$, close to -8.2 deg that optimizes the fit of theory and experiments.

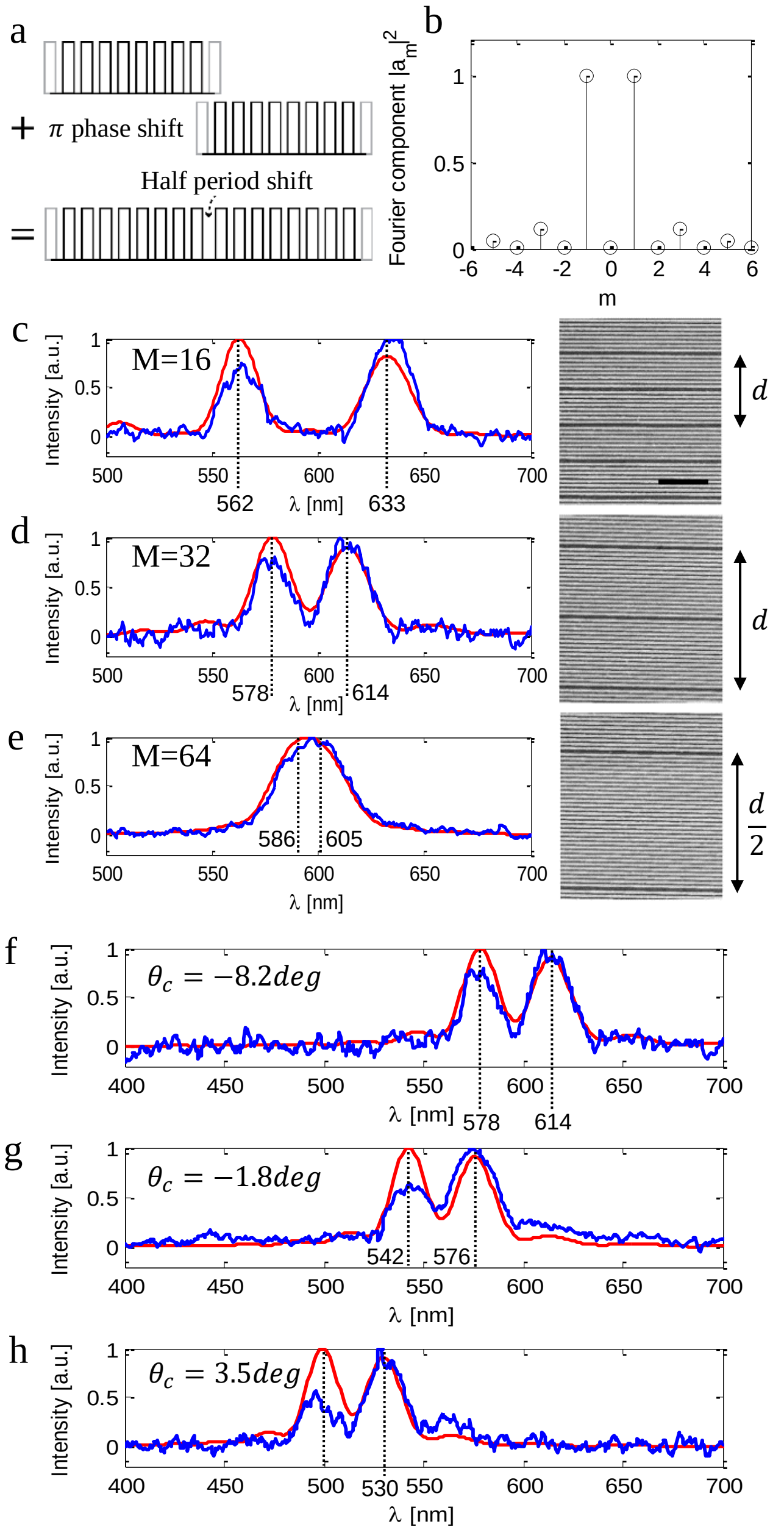

**Figure 2 – Spectral splitting of Smith-Purcell peak.** (a) Schematic demonstration of half-period shift (b) The Fourier components of the phase function, showing two main peaks of the $\pm 1$ orders and a suppressed zero order. (c-h) Theoretical (according to Eq. 4, red) and experimental (blue) Smith Purcell radiation spectrum for twin peak designs. Marked: peak locations according to Eq. 5, for $m = \pm 1$. (c-e) Spectral response for $\theta = -1.8\ deg$, for different M values. $\Lambda$ =400 nm for all three gratings. Insets to (a-c) – Scanning Electron Microscope images of the gratings used for (c-h), scale bar is 5 microns. (f-h) Spectral response for the grating of M=32, for different $\theta$ values.

Fig. 2(f-h) presents the same grating of Fig. 2(d) for different collecting angle $\theta_c$, showing that the main effect of changing the angle of collection is the wavelength translation of the response, much like in regular SPR. Therefore, the far-field distribution of a single wavelength also maintains a twin-peak angular distribution. We note that the difference between the theoretical and the experimental curves is the result of the assumption that $\epsilon_{xn}^{r}(\beta,\omega)$ is constant for different light frequencies. The presented optimized values of $\theta_c$: -8.2 -1.8 and 3.5 degrees for Fig 2(f-h), are well inside the error range ($\pm$1.5 degrees) of the measured values: -7.7, -1.2 and 2.2 degrees respectively.

The finite interaction length limits our spectral control as is apparent from Fig. 2e, where the two peaks merge into a single wider peak. The limited interaction length in our experiments is the result of the angle between the electron beam and the grating, marked as $\varphi$ in Fig. 3(a). Ideally, this angle would be zero, however the width of the electron beam and alignment limitations in our setup prevent us from achieving perfect parallelism. Interestingly, we can also use this angle degree of freedom to our advantage when a wider-spectrum source is desired: increasing $\varphi$ causes the peaks to widen and creates a smoother spectrum due to smaller interaction length. To investigate this further we use a grating design (Fig. 3(b)) that exhibits triple-peak response (the -2, 0 and +2 Fourier coefficients; see [17] for the design of this grating). Fig. 3(c) presents the measured spectrum using the same aperture and $\Lambda$ values as the previous experiment, along with a theoretical curve to which the interaction length was fitted to be 10 $\mu m$ (or 25$\Lambda$). Deliberately increasing the angle $\varphi$ between the electron beam and the grating decreases the interaction length and widens the spectral peaks (Figs.3(d,e)). This confirms the interpretation of SPR as a collective effect to which each unit cell is contributing. Note that the collection angle $\theta_c$ is effectively shifted by $\varphi$ (even when the collection system in the lab remains fixed). The values after optimization for Fig. 3(c-f) were $\theta_c = -7, -6.7, \text{and} -4.8$ deg, and $2x_{max} = 10, 7, \text{and } 4.6$ micron.

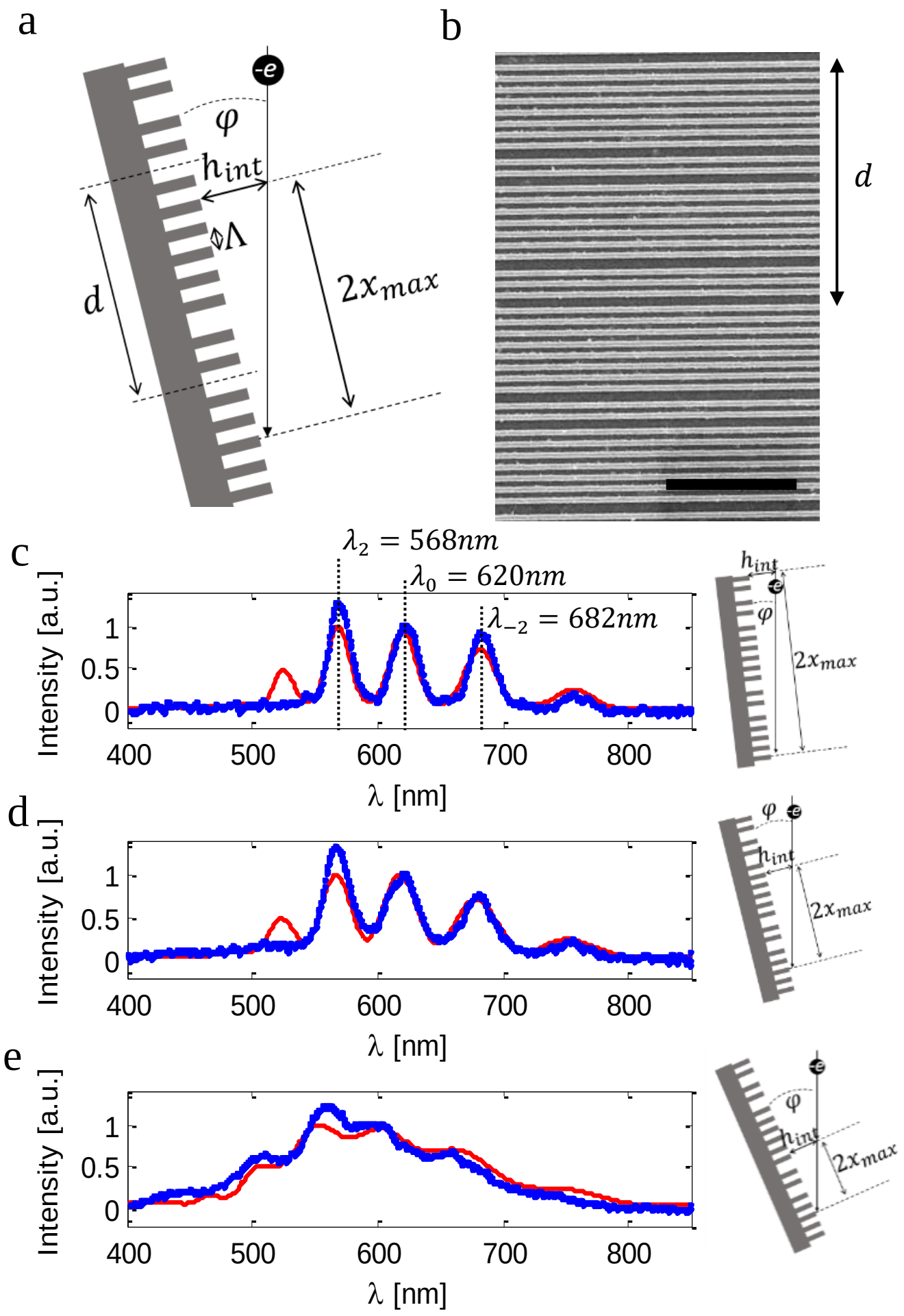

**Figure 3 – Influence of the interaction length on spectral shaping.** (a) – Schematic drawing showing the change of interaction length caused by varying the angle between the electron beam and the grating. (b) – Scanning Electron Microscope Image of grating for 5 peak design. Scale is 5 microns. (c-f) - Theoretical (according to Eq. 4, red) and experimental (blue) Smith Purcell radiation spectrum for three-peak design, for different interaction lengths, changed by altering the angle between the electron beam and the grating. Marked: wavelengths $\lambda_m = \lambda_{m,n=-1}$ according to Eq. 5.

### D. Spatial shaping of SPR

The concepts we outlined above further enable to investigate ways to shape the spatial response of SPR. For instance, we here propose a general design for a "Smith Purcell lens" that will focus the emitted radiation at a predesigned point in $(x, z)$ plane. This way the SPR source will multiplex the functionality of the source and the lens, which for certain spectral ranges (e.g., extreme ultraviolet) may be inefficient or nonexistent. A lens designed for wavelength $\lambda_c$ ($k_c = 2\pi/\lambda_c$) with focal length $f_{len}$, focused to an off-axis point with angle $\zeta$ with z axis, as illustrated in Fig. 4, requires a phase modulation of the form

$$\Delta\psi = k_c\sqrt{f_{len}^2 - x^2} + k_c x \sin\zeta$$

Therefore, the angle of the emitted radiation from point x on the grating should be $sin\theta_n = \frac{1}{k_c}\frac{d(\Delta\psi)}{dx}$. From Eq. (2) we get (for $\phi = \pi/2$):

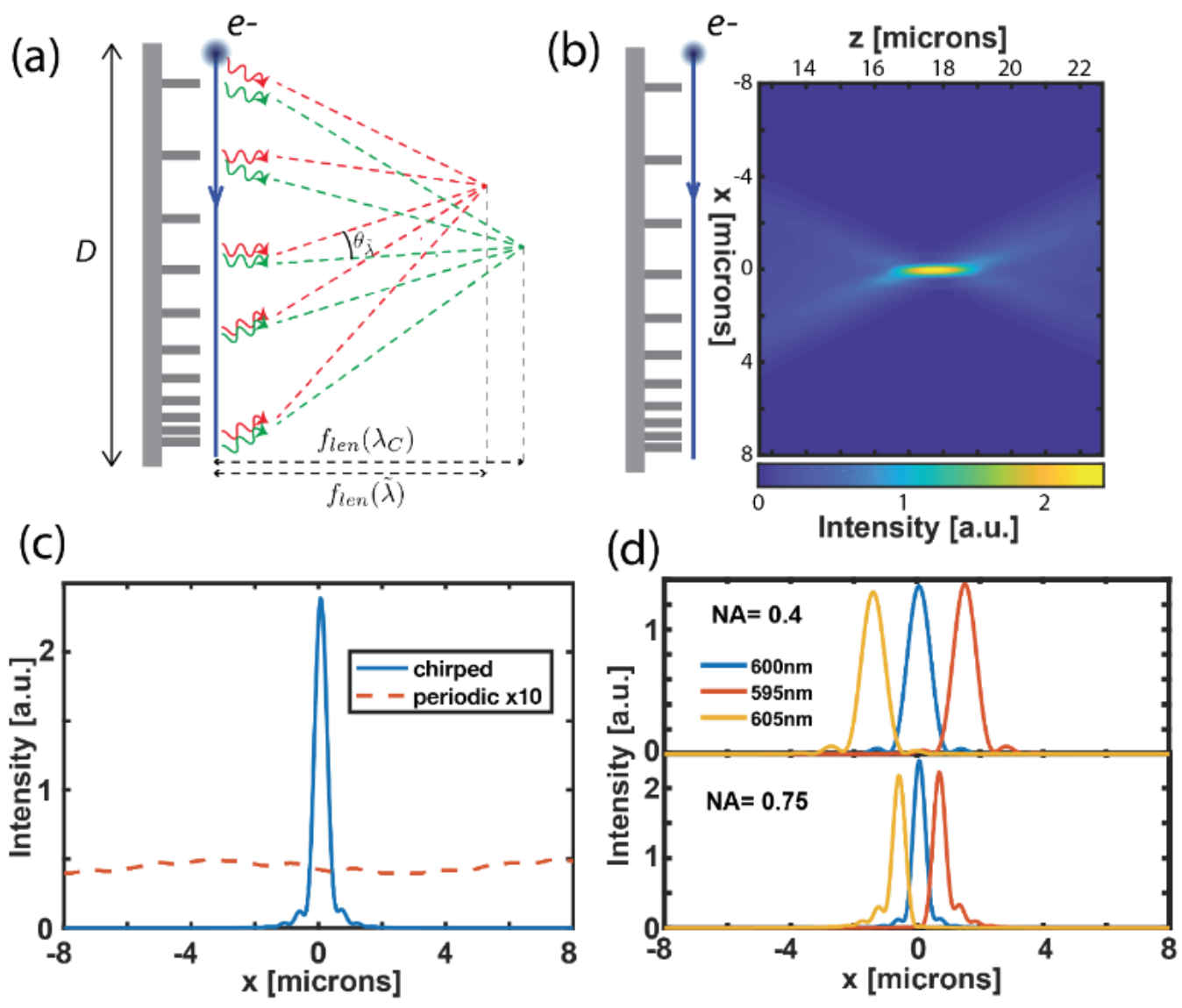

**Figure 4 - Spatial shaping of SPR: design of a Smith-Purcell lens.** (a) Schematic description of Smith-Purcell "cylindrical lens", designed to focus a wavelength $\lambda_c$ at focal distance $f_{len}(\lambda_c)$ and suffering chromatic aberration for other wavelengths. (b) Far-field radiation full-wave simulation as a function of the distance from the chirped grating. The length of the surface is $D$=40µm (giving NA=0.75), the acceleration voltage is 200keV and the focused wavelength $\lambda_C$ = 600nm. (c) Comparison chirped-grating far-field radiation in the focal plane, with a grating of constant period $\Lambda$=163nm (same as the average period of the chirped grating, the intensity being multiplied by 10 to appear on the same scale). (d) Far-field radiation of multiple wavelengths in the focal plane with NA = 0.4 (top), NA = 0.75 (bottom).

$$\Lambda(\mathrm{x}) = \frac{n\lambda_c}{\frac{x}{\sqrt{f_{len}^2 - x^2}} + \sin\zeta - \frac{c}{v_0}} \approx \frac{n\lambda_c}{\frac{x}{f_{len}} + \sin\zeta - \frac{c}{v_0}}$$

This result can also be explained using the shift function $s(x)$ notation [17]. We note that the convergence exists only at x-z plane, and the divergence at the y-z plane is left unchanged. Therefore, this device functions as a cylindrical lens to the emitted radiation as can be seen in our time-domain simulations for a chirped grating focusing SPR at $\lambda_C$ (Fig. 4(c)). If another wavelength $\tilde{\lambda}$ is used other than the design wavelength $\lambda_c$, we'll get (for $x \ll f_{len}$ and $\zeta = 0$):

$$sin\theta = \frac{\tilde{\lambda}n}{\Lambda(\mathrm{x})} + \frac{c}{v_0} = \frac{x}{\frac{\lambda_c}{\tilde{\lambda}} f_{len}} + \left(1 - \frac{\tilde{\lambda}}{\lambda_c}\right)\frac{c}{v_0}$$

Therefore, wavelength $\tilde{\lambda}$ will have a different focal length, given by $\tilde{\mathrm{f}}_{\mathrm{len}} = \frac{\lambda_c}{\tilde{\lambda}}\mathrm{f}_{\mathrm{len}}$. The rightmost term is the tilt expected for wavelength $\tilde{\lambda}$ at x=0, and gives the angle $\theta_{\tilde{\lambda}}$ between the center of the converging cone of each wavelength, $\sin\theta_{\tilde{\lambda}} = \left(1 - \frac{\tilde{\lambda}}{\lambda_c}\right)\frac{c}{v_0}$ (see Fig. 4(a) and (d)). We observe in our FDTD simulations [17] a focusing performance of 11.4% (resp. 17.5%) above diffraction limit for NA = 0.4 (resp. NA = 0.75). We conclude that this cylindrical-lens-like device has strong axial chromatic dispersion, and the expected lateral chromatic dispersion of ordinary SPR. The high chromaticity of this device would be of interest for applications were a small bandwidth is required (deep UV lithography, mask inspection, and biological applications [18–20]) and spectrometry. Even though there has not been, to the best of our knowledge, any experimental demonstration of Smith-Purcell radiation in the deep ultraviolet so far, there are promising theoretical predictions in the literature [21,22], paving the way to an efficient, tunable source in the ultraviolet.

Figure 4(d) shows the highly dispersive nature of a Smith-Purcell lens, which would make difficult the experimental demonstration of its close-to-diffraction-limited behavior with conventional broadband imaging systems (for instance, CCD cameras). This limitation could be circumvented by designing chirped metasurfaces made of resonators with a very high spectral quality factor or plasmon-locked SPR as has been recently proposed [23]. The latter approach relies on phase-matching the plasmon dispersion relation with the electron beam line $\omega = kv$, thus preventing applications to extreme UV regimes. Other dispersion engineering tricks should also be investigated from, for instance, the field of achromatic metasurfaces [24].

### E. Conclusions

In this Letter, we have demonstrated how Smith-Purcell radiation can be shaped to a desired spectral or angular far-field response. Its resolution is limited only by the interaction length of the electrons with the modified grating. By carefully engineering the grating modulation function $s(x)$, and the resulting Fourier components $a_{nm}$ from Eq. (3), it is possible to create a light source with arbitrary spectral and angular shape, at spectral ranges that are unreachable with conventional light sources. While most research in this field so far dealt with simple periodic gratings, for which the spectral and angular response have a single peak (whose location is given by the ordinary Smith Purcell dispersion Eq. (2) [1]), our method enables shaping the radiation spectrum into arbitrary line-shapes, exemplified by our measured dual- and triple-peaked spectra. Furthermore, the method is not limited to periodic phase manipulations, and as we suggested one can use it to create a lens for a specific emitted wavelength, converging the beam in x-z plane. This might be useful for spectral ranges where lenses are hard to achieve, such as the UV or x-ray ranges [5].

We note that although the interaction length seems to impose a limitation on the widths of the peaks, the concept of super-oscillations [25,26] might be used together with our method to achieve peaks at arbitrary small widths (with penalty to the emitted intensity). The concepts of holography [27] can be used to design the structuring function $s(x)$ for shaping the complex Smith-Purcell emitted waveform (as was already demonstrated for transition radiation [28]). This can be achieved for example by varying other parameters of the grating structure such as the duty cycle, the ridge height, or the profile shape of the periods. Furthermore, in our theoretical analysis the value of $\epsilon_{xn}^{r}(\beta,\omega)$ is assumed constant, which explains the deviation of the peaks height from the theoretical curves. Therefore, this approach might be used to measure the function $\epsilon_{xn}^{r}(\beta,\omega)$ at a single spectral measurement.

The authors would like to thank Dr. Yigal Lilach for assistance in fabrication, and Prof. Avi Gover for stimulating discussions. This work was supported by DIP, the German-Israeli Project cooperation and by the Israel Science Foundation, Grant No. 1310/13, and by the U. S. Army Research Office through the Institute for Soldier Nanotechnologies (Contract No. W911NF-13-D-0001). I.K. was supported by the FP7-Marie Curie International Outgoing Fellowship (IOF) under Grant 328853-MC-BSiCS.

Supplemental Material: **Spectral and spatial shaping of Smith Purcell Radiation**

Roei Remez[1]*, Niv Shapira[1], Charles Roques-Carmes[2], Romain Tirole[2,3], Yi Yang[2], Yossi Lereah[1], Marin Soljačić[2], Ido Kaminer[2,4] and Ady Arie[1]
[1]School of Electrical Engineering, Fleischman Faculty of Engineering, Tel Aviv University, Tel Aviv, Israel.
[2]Research Lab of Electronics, MIT, Cambridge, MA 02139, USA
[3]Department of Physics, Faculty of Natural Sciences, Imperial College London, London SW7 2AZ, UK
[4]Department of Electrical Engineering, Technion – Israel Institute of Technology, Haifa 32000, Israel
*Correspondence to: roei.remez@gmail.com.

## A - Detailed theory of Smith-Purcell for complex periodic gratings

This section develops the SPR theory following the work of Van Den Berg [1], and applying it to the case of complex periodic gratings as in Figs. 2 and 3 of the manuscript. We start with writing the expression of the field emitted from an infinite metallic periodic grating, positioned at (*x*, *y*) plane with the rulings parallel to the *y* direction, and where the electron is passing close to the grating with velocity $v_0\hat{x}$ [1]:

$$E_x(x,y,z,t) = \frac{1}{2\pi^2} Re\left[\int_0^\infty d\omega \int_{-\infty}^{\infty} d\beta \sum_{n=-\infty}^{\infty} \epsilon_{xn}^r(\beta,\omega) e^{i\alpha_n x + i\beta y + i\gamma_n z - i\omega t}\right] \quad (1)$$

Where:

$$\alpha_n = \frac{\omega}{v_0} + \frac{2\pi n}{\Lambda} = k_0 \sin\phi \sin\theta_n$$

$$\beta = k_0 \cos(\phi)$$

$$\gamma_n = \sqrt{k_0^2 - \beta^2 - \alpha_n^2} = k_0 \sin\phi \cos\theta_n$$

Where $k_0 = \frac{\omega}{c} = \frac{2\pi}{\lambda_0}$, $\lambda_0$ and $\omega$ are the vacuum wavelength of light and its angular frequency, *c* is the velocity of light, $\phi$ is the angle of the wave vector with y axis and $\theta_n$ is the angle of the projected *k* vector on the (*x*, *z*) plane with *z* axis. The discrete summation over *n* originates from the different diffraction orders of the grating, with corresponding spatial frequency components $\alpha_n$ and $\gamma_n$. Although the summation is on all values of *n*, we note that for high diffraction orders $\gamma_n$ becomes imaginary and the wave is therefore evanescent. From the equation for $\alpha_n$ we can formulate the dispersion equation:

$$\lambda_0 = \frac{\Lambda}{n}\left(\sin\phi \sin\theta_n - \frac{c}{v_0}\right) \quad (2)$$

Which is equivalent to the dispersion relation from the first paper by Smith and Purcell [2]. Eq. (1) represents decomposition into plane waves in direction $(\theta_n, \phi)$. Defining $E_{xn}$ as the electric field of the $n^{\text{th}}$ diffraction order,

$$E_{xn} = \frac{1}{2\pi^2}\int_0^\infty d\omega \int_{-\infty}^{\infty} d\beta\, \epsilon_{xn}^r(\beta,\omega) e^{i\alpha_n x + i\beta y + i\gamma_n z - i\omega t},$$

we can approximate the electric field emitted in the case of a structured grating (SG) as:

$$E_x^{SG}(\mathrm{x},\mathrm{y},0,\mathrm{t}) = \mathrm{rect}\left(\frac{x - x_0}{2x_{max}}\right) \sum_{n=-\infty}^{\infty} f_n(x) E_{xn}(x,y,0,t) \quad (3)$$

Where $2x_{max}$ is the interaction length of the electron with the grating, $x_0$ is the center of the interaction with respect to an arbitrary point at the grating, and $f_n(x)$ denotes the effect of the deviation from a periodic grating on the electric field of the $n^{\text{th}}$ diffraction order at z=0. We note that in a single SPR experiment, usually many values of $x_0$ are measured simultaneously, since the non-zero width of the electron beam gives different interaction area for each electron.

For simplicity, we further limit our derivation for the case of constant duty cycle (ridge to period ratio) of 50%, which results with phase-only changes for the electric field. We can define a function $s(x)$ that expresses the structuring of the grating (the results below can be directly generalized to any duty cycle and even arbitrary grating shapes). A shift of $s(x)\Lambda$ (where $0 \le s(x) \le 1$) will result in phase shift $\Delta\psi$ for the electric field at *z=0*, given by:

$$\Delta\psi = 2\pi s(x)n$$

where $n$ is the SPR diffraction order (see the next section for Huygens construction of this expression). Therefore, $f_n(x) = e^{i2\pi s(x)n}$. Here, the function $s(x)$, and therefore $f_n(x)$, are assumed periodic, however this does not have to be the case, and some interesting results can be achieved with non-periodic variations of the grating.

The electric field at the grating plane can be rewritten as a far-field expression, defining the far-field angular distribution $\eta_x^r(\alpha,\beta,\omega)$:

$$E_x^{PG}(x,y,0,t) \equiv \frac{1}{2\pi^2}\int_0^\infty d\omega \int_{-\infty}^\infty d\beta \int_{-\infty}^\infty d\alpha\, \eta_x^r(\alpha,\beta,\omega)e^{i\alpha x+i\beta y-i\omega t} \qquad (4)$$

Comparing Eq. (1) and (4) and substituting Eq. (3), we get

$$\eta_x^r(\alpha,\beta,\omega) = 2x_{max}\sum_{n=-\infty}^{\infty} e^{-i\alpha x_0}\epsilon_{xn}^r(\beta,\omega)\,\mathrm{sinc}((\alpha-\alpha_n(\omega))x_{\max}) \otimes \mathrm{FT}\{f_n(x)\}$$

Where $\mathrm{FT}\{f_n(x)\}$ is the Fourier transform of $f_n(x)$, and $\otimes$ denotes convolution. Since $f_n(x)$ is periodic, we can write its Fourier transform as a Fourier series $FT\{f_n(x)\} = \sum_{\mathrm{m}=-\infty}^{\infty} a_{mn}\delta\left(\alpha - \frac{2\pi m}{d}\right)$, where $d$ is the period of $f_n(x)$ and $a_{mn}$ are Fourier coefficients. Finally, the far-field expression is given by

$$\eta_x^r(\alpha,\beta,\omega) = 2x_{\max}\sum_n \epsilon_{xn}^r(\beta,\omega)\sum_m a_{mn}e^{-i\left(\alpha-\frac{2\pi m}{d}\right)x_0}\mathrm{sinc}\left(\left(\alpha-\frac{2\pi m}{d}-\alpha_n(\omega)\right)x_{\max}\right) \quad (5)$$

## B - The effect of period shift on the phase of the field

Here, we show the effect of period shift $s(x)$ on the phase of the $n$'th order of the electric field at the plane of the grating. Given a grating with period $\Lambda$, and a shift of $s\Lambda$ from a specific point at the grating, we'll show that the phase shift $\Delta\psi$ of the electric field is constant for all wavelengths and equal to:

$$\Delta\psi = 2\pi ns$$

Where $n$ is the diffraction order. We'll use Huygens construction method to show both the above equation and the ordinary Smith Purcell dispersion equation.

Let us define $t_0$ as the time it takes the electron to travel one grating period $\Lambda$, that is, $t_0 = \frac{\Lambda}{\beta c}$. The electron passing point A (towards B) emit light with phase 0 that propagates after time $t_0$ to point c, $AC = t_0 c$. In the same time, the electron propagates to point B, and when reaching it (after time $t_0$), emit light with phase 0. The phase difference between points B and C is zero, and therefore the distance BD is the wavelength of the light times an integer: $\lambda n = BD = sin\theta AB + AC = \Lambda sin\theta + \frac{\Lambda}{\beta}$. From this we get the Smith Purcell dispersion equation: $sin\theta = \frac{\lambda n}{\Lambda} - \frac{1}{\beta}$.
What happens if the grating is shifted at point E, by $s\Lambda$? Now the distance between two zero phase lines is EF. Since from this point on the period remains $\Lambda$, the angle $\theta$ remains constant. The phase jump is therefore $\Delta\Psi = 2\pi\left(\frac{\mathrm{EF}}{\lambda}\right) = 2\pi\left(\frac{\mathrm{EF}}{\lambda}\right)$. Since $EB = s\Lambda sin\theta + \frac{s\Lambda}{\beta}$, and $\lambda = \frac{1}{n}(\Lambda sin\theta + \frac{\Lambda}{\beta})$ we get $\Delta\Psi = 2\pi ns$. This expression is correct for all wavelengths and angles.

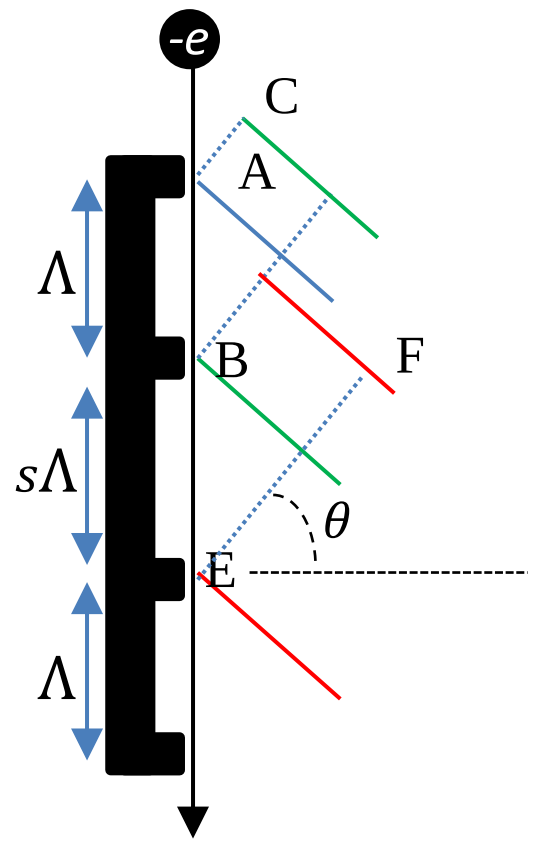

**Figure S1** – Huygens reconstruction for the emitted radiation in the case of period shift by factor s.

## C - Experimental details

The gratings for the experiments in this Paper were fabricated using E beam lithography on a 500 micron BK7 substrate, and coated with 200nm of silver. The gratings were placed inside the viewing chamber of an FEI-Titan 200KeV electron microscope, and were aligned to be almost parallel to the electron beam. The optical setup outside the microscope's viewing chamber consisted of two 2" doublet lenses, chromatically corrected for the measured spectral range, and a collecting fiber, connected to an Avantes ULS2048L spectrometer with a spectral resolution of 1.4 nm. A circular aperture of 13mm was used to limit the angular acceptance of the light collection system in all the experiments in this Letter.

## D - Design of 3-peak grating

This section shows the design of the complex periodic grating used in Fig. 3 to generate a triple-peak SPR. For the design of triple peak grating (see also Fig. S2), we used the following $s(x)$ function (over one supercell period):

$$s(x) = \begin{cases} 0 & 0 < x \leq \frac{8}{20}d \quad or \quad \frac{d}{2} < x \leq \frac{17d}{20} \\ 0.5 & \frac{8}{20}d < x \leq \frac{d}{2} \quad or \quad \frac{17d}{20} < x \leq d \end{cases}$$

This grating exhibit three peaks (corresponding to the -2, 0 and +2 Fourier coefficients) with approximately equal amplitude, and additional two lower peaks (corresponding to the -4 and +4 Fourier coefficients), as shown in Fig 3(b). All odd components of the Fourier sum are suppressed.

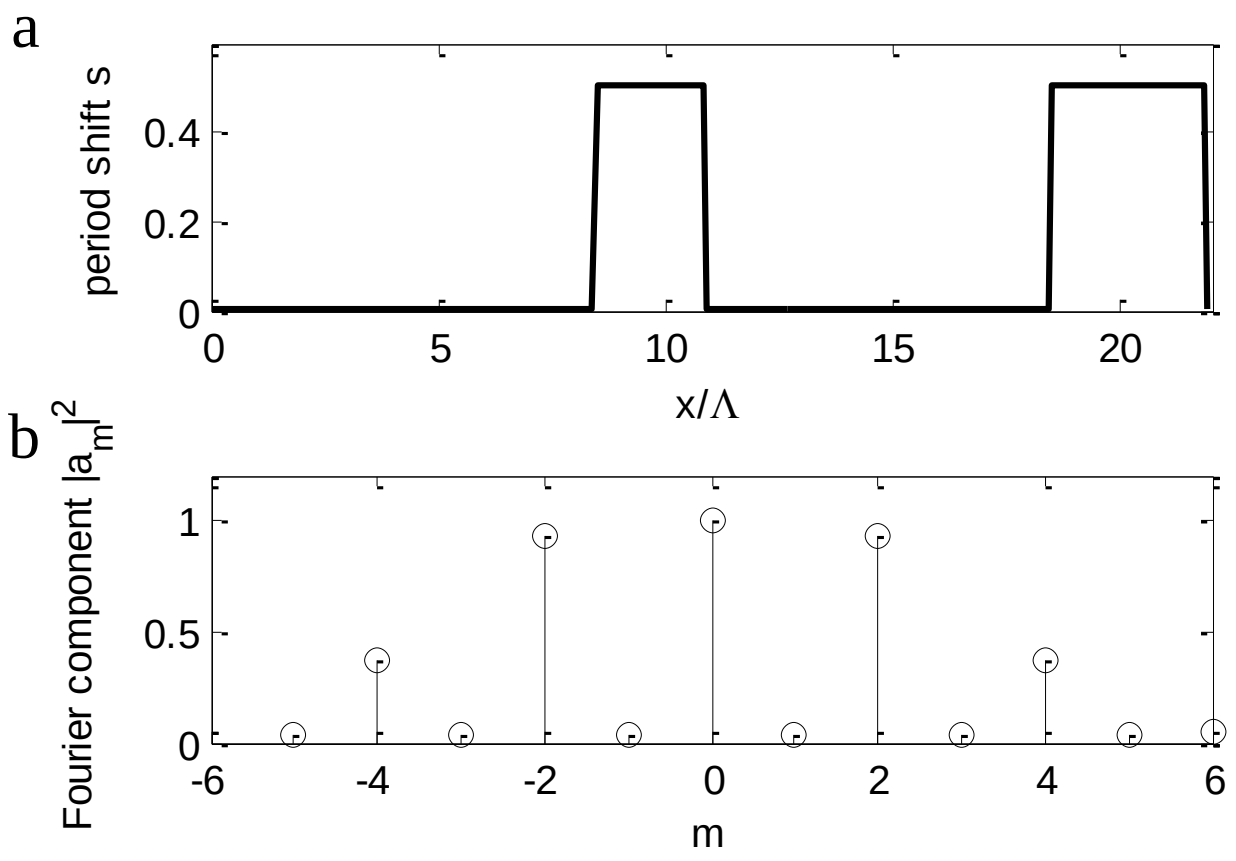

**Figure S2 – Design of 3-peak grating.** (a) – period shift function $s(x)$ as a function of number of periods $x/\Lambda$. (b) – Fourier sum components $|a_m|^2$ for the grating shown in (a), showing 5 peaks with the central three peaks with approximately equal amplitude and suppressed odd orders.

## E - Lens grating described by period shift function $s(x)$

In this section we show that the shape of the lens grating can be formulated using the functions $f(x)$ and $s(x)$, even though they were initially formulated for a constant grating period $\Lambda$. Identifying $\Lambda(x) = \left(\frac{ds(x)}{dx}\Lambda + 1\right)\Lambda$, and limiting this demonstration to $\zeta = 0$ and $NA << 1$, we get:

$$s(x) = \frac{1}{\Lambda^2}\int (\Lambda(x) - \Lambda)dx = \frac{1}{\Lambda}\int \frac{1}{1 - \frac{v_0 x}{c f_{len}}} dx - \frac{x}{\Lambda} = -\frac{c f_{len}}{v_0}\frac{1}{\Lambda}\ln\left(1 - \frac{v_0 x}{c f_{len}}\right) - \frac{x}{\Lambda}$$

$$\approx \frac{c f_{len}}{v_0}\left(\frac{v_0 x}{\Lambda c f_{len}} + \frac{1}{2\Lambda}\left(\frac{v_0 x}{c f_{len}}\right)^2\right) - \frac{x}{\Lambda} = \left(\frac{x}{\Lambda} + \frac{1}{2\Lambda}\frac{v_0 x^2}{c f_{len}}\right) - \frac{x}{\Lambda}$$

$$= \frac{1}{2\Lambda}\frac{v_0 x^2}{c f_{len}}$$

$$\Delta\psi = 2\pi n s(x) = 2\pi n\left(\frac{1}{2\Lambda}\frac{v_0 x^2}{c f_{len}}\right) = 2\pi\left(\frac{1}{2\lambda_c}\frac{x^2}{f_{len}}\right) = \frac{2\pi}{\lambda}\left(\frac{x^2}{2\frac{\lambda_c}{\lambda} f_{len}}\right)$$

Where we've used the approximation $\frac{v_0 D}{2c f_{len}} \ll 1$ and $\Lambda(x = 0) \equiv \Lambda = -n\lambda_c v_0/c$.

## F – Equivalence of the theory with a dipole array model

We have developed an additional theoretical method to predict the SPR from complex gratings, and it matches with the results of the method presented in the main text. We model the electric field induced by the flight of the electron by an array of dipole sources located at the positions of the ridges of the complex periodic grating. The relative phase between the dipoles is set by the time of arrival of the electron to a point above that ridge (hence the relative phase is the only parameter that depends on the electron velocity). This simple dipole array model allows to predict a far-field spectral radiation pattern very similar to our experimental results, for example in the case of the first-order splitting design (Fig. S3). The total number of dipoles, corresponds to the number of unit cells that are excited in a coherent manner, is extracted from experimental data. The angle of observation is known from the geometry of the system and can also be fitted to obtain the same center wavelength. The spacing for different values of M matches the experimental data.

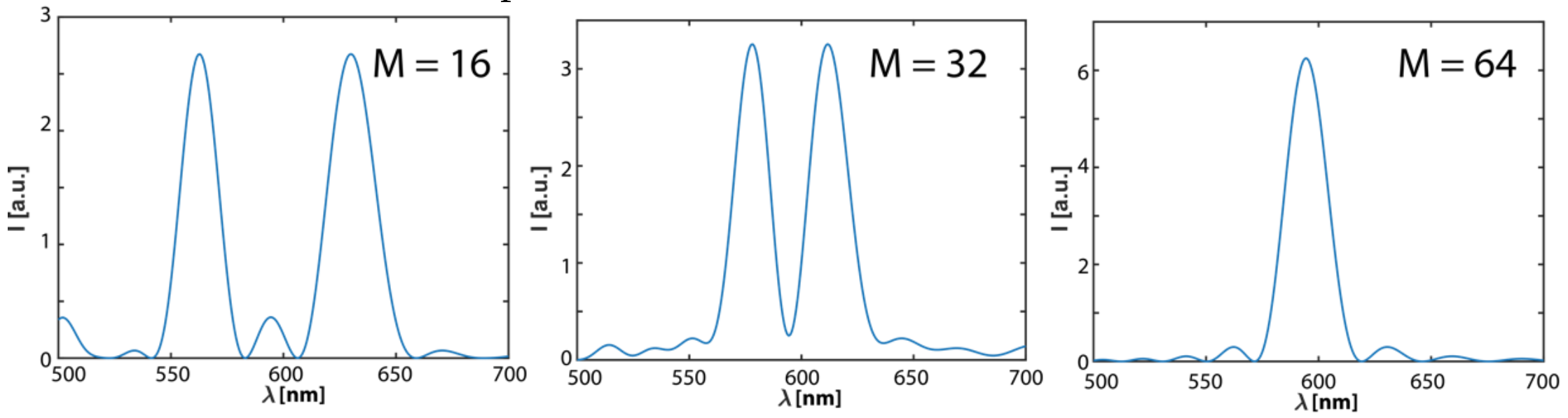

**Figure S3 – SPR spectral peak splitting predicted by the dipole array model.** Far-field radiation pattern from an array of dipoles as in the complex periodic gratings of Fig. 2, exhibiting SPR first-order splitting, for different M.

## G – Smith-Purcell lens – Experimental limitations

In this section, we discuss the dispersive nature of a Smith-Purcell lens and the influence of the electron beam interaction length on diffraction-limited behavior.

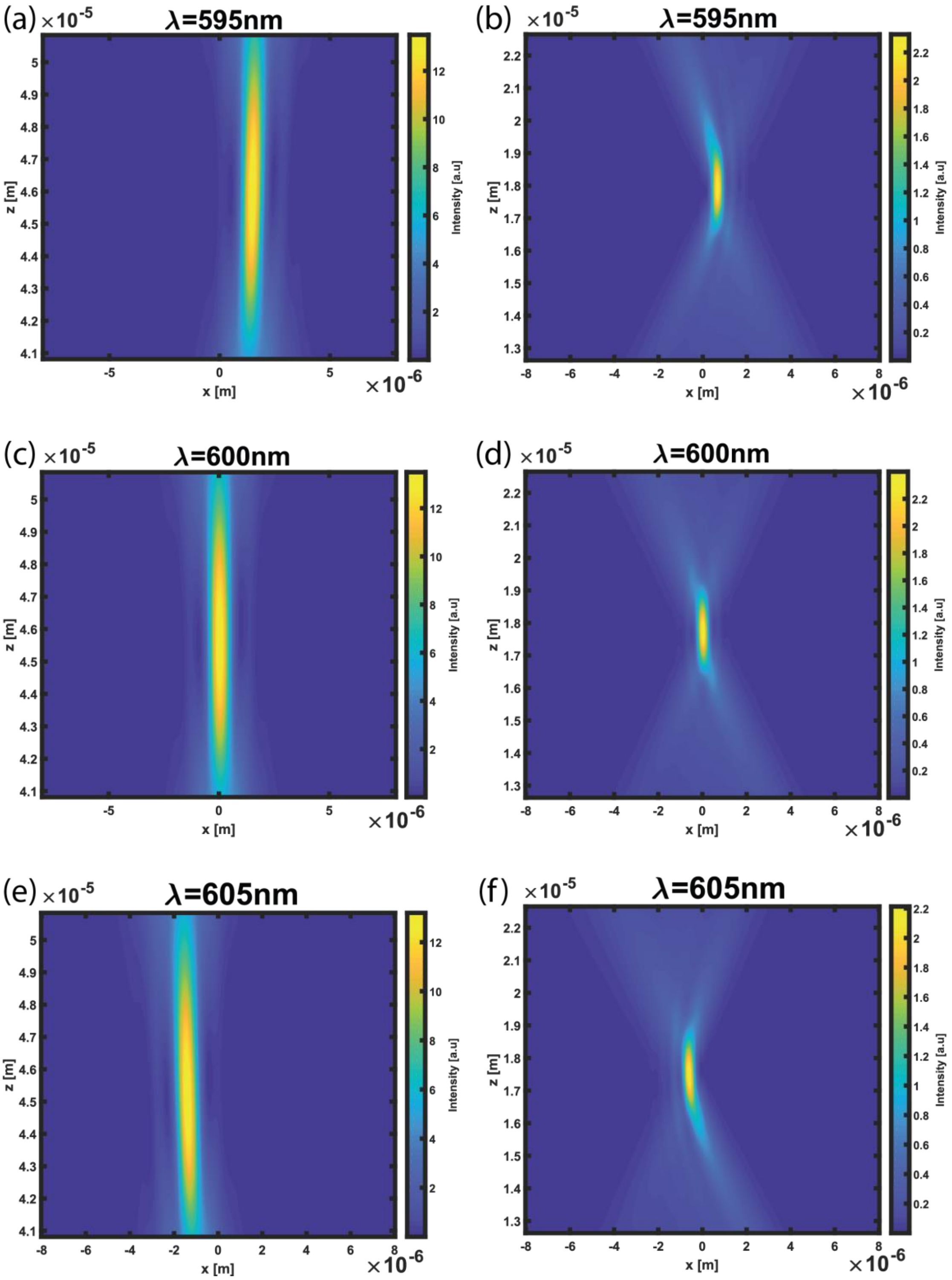

**Figure S4 – Chromatic aberrations of Smith-Purcell lens.** Far-field distribution in the (x,z) plane for a Smith-Purcell lens of diameter D=40$\mu$m, NA = 0.4 ((a)-(c)-(e)) or NA = ((b)-(d)-(f)) at different wavelengths around the design wavelength $\lambda_c = 600$nm.

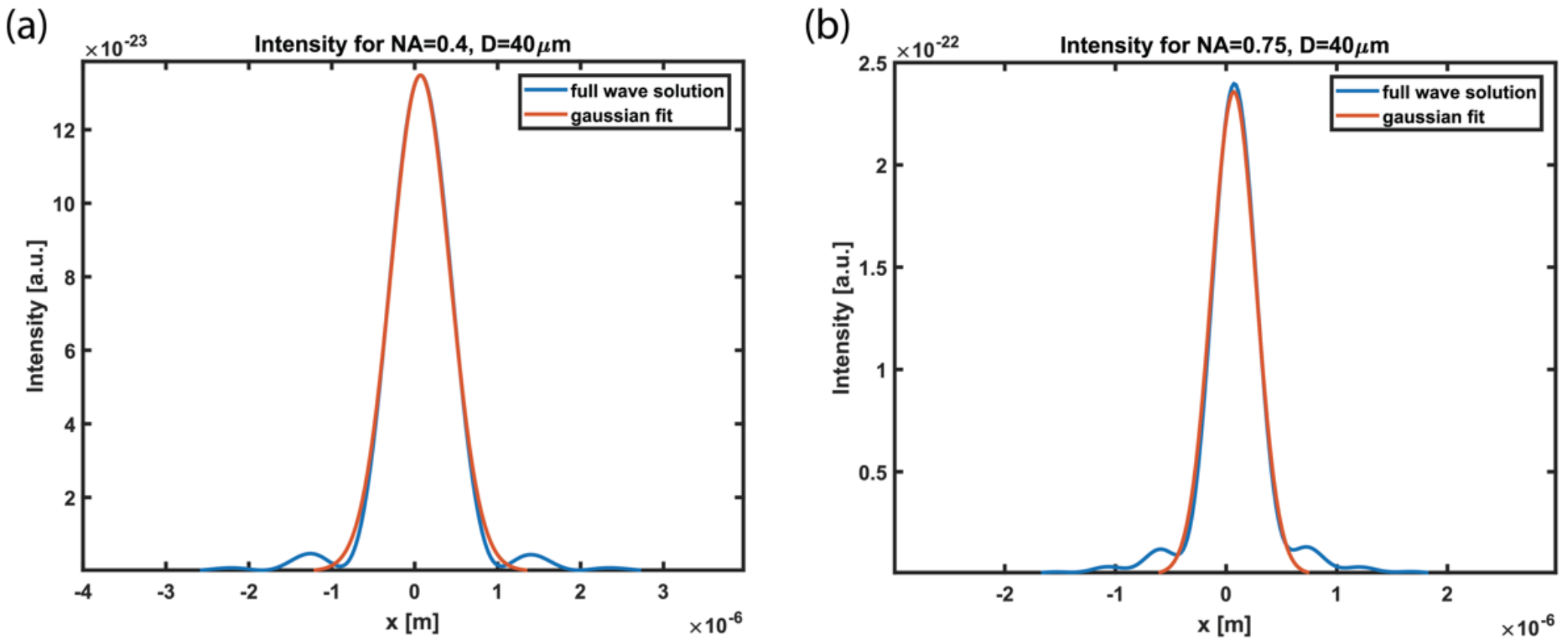

**Figure S5 - Intensity profile in the focal plane for (a) NA = 0.4 and (b) NA 0.75.** In both cases, the intensity profile is fitted with a gaussian in order to estimate its FWHM and compare it with a diffraction-limited behavior.

Figure S4 shows the highly dispersive nature of a Smith-Purcell lens, which would make difficult the experimental demonstration of its close-to-diffraction-limited behavior with conventional broadband imaging systems (for instance, CCD cameras). This first limitation could be circumvented by designing chirped metasurfaces made of resonators with a very high spectral quality factor or by using already explored dispersion engineering tricks from, for instance, the field of achromatic metasurfaces [3]

Moreover, the interaction length in the case of a chirped grating would influence its diffraction-limited performance. If the interaction length includes the entire structure, we can expect spot size cross section similar to the one of a diffraction limited cylindrical lens, exhibiting spot of diameter $\sim \frac{\lambda_C}{2\text{NA}}$, $NA = \sin\left(atan\left(\frac{D}{2f_{len}}\right)\right)$ where $D$ is the length of the grating. We observe in our simulations a focusing performance of 11.4% (resp. 17.5%) above diffraction limit for NA = 0.4 (resp. NA = 0.75). However, an interaction length $2x_{\max}$ smaller than $D$, while still assuming the electron beam covers the entire length $D$, will require incoherent summation over all possible locations of the interaction over the grating. We conclude that the expected spot diameter of such a lens (or, more correctly, the line width) will be limited by an effective $NA = \sin\left(\text{atan}\left(\frac{\min(D, 2x_{max})}{2f_{len}}\right)\right)$. An experimental observation of this effect would require an interaction length and numerical aperture large enough for the focusing effect to become apparent. Also, the large chromaticity of this device would significantly broaden its focal spot (see Fig. 4(d)), assuming a detector or camera would record some bandwidth around the design wavelength. These experimental challenges still need to be addressed.

## H – Time-domain simulation setup

This section explains the time-domain simulation used in Fig.4b. The electron beam can be represented as a time-dependent propagating point-electron:

$$J(\vec{r}, t) = -e\, v\, \delta(x - vt)\delta(y - y_0)\delta(z - z_0)$$

and the associated polarization $P(\vec{r}, \omega) = i\frac{e}{\omega}\exp(-i\frac{\omega x}{v})\ \delta(y - y_0)\delta(z - z_0)\ \vec{x}$. Time-domain simulations are run using the commercial FDTD software Lumerical. We model the electron beam by a set of closely spaced dipole sources, in order to induce a polarization similar to $P(\vec{r}, \omega)$.